\documentclass[twocolumn,twoside,slac_two]{revtex4}
\usepackage{graphicx}
\usepackage{fancyhdr}
\begin{document}

\title{Radio-loud AGN contribution to the Extragalactic Gamma-Ray Background}

\author{Debbijoy Bhattacharya}
\affiliation{Inter-University Centre for Astronomy\& Astrophysics,
  Pune-411007, India}
\author{M. Errando, R. Mukherjee}
\affiliation{Barnard College,
  Columbia University, New York, NY 10027, USA}
 \author{M. B$\ddot{o}$ttcher}
\affiliation{Department of Physics and Astronomy, Ohio University, Athens, OH 45701, USA}
%\author{R. Mukherjee}
%\affiliation{Barnard College,
%  Columbia University, New York, USA}
\author{P. Sreekumar}
\affiliation{Space Astronomy Group, ISRO Satellite
  Centre, Bangalore, India}
 \author{R. Misra}
\affiliation{Inter-University Centre for Astronomy\& Astrophysics,
  Pune, India}
\author{P. Coppi}
\affiliation{Department of Astronomy, Yale University, P.O. Box 208101, New Haven, CT 06520, USA}
%\author{M. Boettcher$^{2}$, P. Coppi$^{3}$, M. Errando$^{4}$,  R. Misra$^{1}$, R. Mukherjee$^{4}$, P. Sreekumar$^{5}$}
%\affiliation{1. Inter-University Center for Astronomy\& Astrophysics,
%  Pune, India, 2. Ohio University, USA, 3. Yale University, USA, 4. Barnard College,
%  Columbia University, USA, 5. Space Astronomy Group, ISRO Satellite
%  Centre, Bangalore India}
%
%\author{M. Boettcher}
%\affiliation{FNAL, Batavia, IL 60510, USA}

\begin{abstract}
Origin of extragalactic $\gamma$-ray background (EGRB) is still a matter of 
debate. EGRB can either have truly diffuse or unresolved discrete point 
sources origin. Majority of the Fermi and EGRET detected identified sources 
are blazar. So, they are expected to be a significant contributors to the EGRB. 
In order to estimate their contribution to the EGRB one needs to construct their 
luminosity functions. Here, we examine and construct the evolution and luminosity 
function of blazars using first LAT AGN catalog. We consider both pure luminosity 
and pure density evolution models for flat spectrum radio quasars (FSRQs). We also describe the methodology to 
estimate the contribution from misaligned blazars to the EGRB.
\end{abstract}

\maketitle

\thispagestyle{fancy}

\section{Introduction}
%The first evidence for an isotropic diffuse $\gamma$-ray emission, probably of extragalactic origin 
%having relatively softer spectrum than that of galactic component was shown by OSO-3 observations \citep{kraus2}. 
%\citet{thompson1982} carried out the final derivation 
%of the spectrum of Extragalactic Gamma-ray Background (EGRB) above $30$ MeV (single power law of spectral index $\sim 2.35$) 
%from SAS-2 satellite data. After the launch of Compton Gamma-Ray Observatory (CGRO) 
%in 1991 \citep{kanbach1988}, \citet{sreekumar1998} derived the EGRB from EGRET data. The differential photon spectrum of EGRB 
%of energy range $30$ MeV - $10$ GeV, averaged over full sky, is well fitted by a power law of index $-(2.10\pm 0.03)$. 
%
%The integrated flux above $100$ MeV is  $(1.45\pm 0.05)\times 10^{-5}$ photon cm$^{-2}$ s$^{-1}$ sr$^{-1}$. 
%Following a different approach, \citet{strong2004a,strong2004b} found a much softer EGRB spectrum 
%which could not be fitted with a single power law and shows some positive curvature beyond 1 GeV. 
%
Observational $\gamma$-ray astronomy got a massive boost after the launch of the Fermi Gamma-ray Space Telescope (Fermi) 
on 11th June, 2008.  The Large Area Telescope (LAT)
onboard Fermi provides an increase in sensitivity by more
than an order of magnitude over EGRET and AGILE. 

The origin of extragalactic $\gamma$-ray background (EGRB) is one of the fundamental unsolved problem in astrophysics. 
EGRB can arise from some diffuse processes like black hole evaporation, large scale structure formation, 
matter-antimatter annihilation, etc (e.g., \cite{loeb2000,gabici2003,cohenrujulaglashow1998}). 
Alternatively, due to the limited instrument sensitivity, unresolved $\gamma$-ray sources 
could contribute significantly to the observed EGRB. 

Recently, the EGRB has been rederived from Fermi Large Area Telescope (LAT) data \citep{abdo2010b}. 
EGRB is consistent with 
a featureless power law of index $(2.41\pm 0.05)$. The integrated flux above $100$ MeV is 
$(1.03\pm 0.17)\times 10^{-5}$ photon cm$^{-2}$ s$^{-1}$ sr$^{-1}$. 
%Further investigation showed that the discrepancy cannot be attributed to a lower threshold for resolving 
%point sources.

\par Fermi already detected $\sim 1500$ sources in 11 months of observation (First Fermi catalog; \cite{abdo2010c}) 
and is expected to detect many more. In contrast to the less than $100$ active galactic nuclei (AGNs) 
in the EGRET catalog, the first Fermi-LAT AGN catalog \citep{abdo2010a} 
consist of $\sim 600$ blazars. Fermi also detected 11 non-blazar AGN \citep{abdo2010d}, three normal 
(M $31$, LMC and Milky Way) and two starburst galaxies (M $82$ and NGC $253$).
Thus, Fermi provides an excellent opportunity to understand the origin of EGRB.
% Potential source classes that can contribute to the background are 
%blazars, misaligned blazars, normal and starburst galaxies, cluster of galaxies and $\gamma$-ray bursts (GRB).

Since the dominant class of identified EGRET and Fermi sources are blazars, 
they are expected to contribute to the EGRB. The contribution from blazars to the EGRB is 
estimated using two different approaches:
\newline (a) Considering a relationship between the blazar $\gamma$-ray luminosity 
with their luminosity at some other wavelength, their $\gamma$-ray luminosity function is assumed to be  
the scaled luminosity function at that wavelength (e.g., \cite{stecker1996,stecker1993,narumoto1,
narumoto2,abazajian2010,inoue2011,li2011,cavadini2011,stecker2011}).
%Many published studies of the EGRB calculated the
%contribution from AGNs considering either a radio/$\gamma$-ray correlation or an X-ray/$\gamma$-ray
%correlation to find the $\gamma$-ray luminosity function of blazars (e.g., [24], [25], [58], [59], [26], [27],[28]).
\newline (b) Constructing the $\gamma$-ray luminosity function directly from observed $\gamma$-ray blazars (e.g., \cite{chiang1995,chiang1998,deb1,deb2}).

%Stecker and his collaborators considered a correlation between $\gamma$-ray and radio luminosity of 
%blazar \citep{stecker1993,stecker1996,stecker2011}. They claimed that 
%blazars are the prime contributor to the EGRB. \citet{narumoto1,narumoto2} considered a correlation between $\gamma$-ray and X-ray emission from 
%blazars and found a $25-50$\% contribution from unresolved blazars to the EGRB. 
%\citet{chiang1995} and \citet{chiang1998} constructed the luminosity function of
%blazars from EGRET-detected sources only and they estimated the unresolved blazar contribution is $\sim 20$\%. 
In an earlier work \citep{deb1,deb2} following a similar approach of \citet{chiang1998} but with almost twice the number of sources, 
we constructed the $\gamma$-ray luminosity functions of flat spectrum radio quasars (FSRQs) and BL Lacs from EGRET-detected 
sources only. We found strong positive evolution for FSRQs, 
while BL Lacs did not show any significant evolution. The maximum contribution from blazars to EGRB was found to be $\sim 20$\%. 
However, due to the limited number of sources in the EGRET catalog, the luminosity function and hence, 
the contribution from blazars was not well constrained. 
\citet{mucke} estimated the contribution from FSRQs and BL Lacs to the EGRB in the context of AGN unification paradigm, 
and found that unresolved blazars (both ``aligned'' and ``misaligned'') contribute $20\%-40\%$ to the EGRB. 
In a similar approach, \citet{dermer2007} estimated that FSRQs and BL Lacs contribute $\sim 10\%-15$\% 
and $2\%-3\%$ at $1$ GeV respectively. 
From the first three months of Fermi observation, Abdo et al.~\cite{abdo2009a} found that FSRQs exhibit strong evolution while BL Lacs 
show no evolution. 
Using the 1st Fermi catalog, \citet{abdo2010e} found from the 
source count distribution that the point source contribution is $\sim 16 \%$ of the EGRB.

%Our earlier works involved construction of luminosity functions of flat 
%spectrum radio quasars (FSRQs) and BL Lacs from pre-Fermi (EGRET) data and estimation of its 
%contribution to the EGRB \citep{deb1,deb2}.  
%However, due to the limited number of sources in the EGRET catalog, the luminosity function and hence, 
%the contribution from blazars was not well constrained. 
%Clearly, this is an opportune time to re-visit this scientific investigation.

\par  
%The luminosity function of BL Lacs determined from Fermi data is
%harder than that calculated for EGRET. this could arise from many factors. Being strongly time-variable sources, 
%one needs to derive the time-averaged source characteristics to address contributions to the background. 
%This consideration was incorporated into the
%characterization of EGRET-detected sources by taking average over the full mission in 
%our early work \citep{deb1}. However, from that analysis the break luminosity and the low 
%luminosity end index could not be determined.  With Fermi detection of many 
%more sources, this situation improves significantly. 

Here, we construct the luminosity function of FSRQs and BL Lacs following the methodology 
described in \citet{deb1}. Utilizing the Fermi observed sources we construct a complete 
source list, apply the ${V}/{V_{max}}$ method to search for the presence of evolution, 
find parameters by modified ${V}/{V_{max}}$ method, and finally construct the luminosity function 
using ${1}/{V_{max}}$ method and maximum likelihood estimator. 
%Since the luminosity function constructed by $\frac{1}{V_{max}}$ can have some bias specially at low luminosity, 
%we use the maximum likelihood estimator to find the parameter value of the luminosity function. 

%Here, we study the $\gamma$-ray properties of blazar source class 
%and use this to better address the unresolved-source contribution to the EGRB. Using the first Fermi AGN catalog, 
%we investigate the evolutionary properties of FSRQs and BL Lacs, construct luminosity functions and estimate 
%their contribution to the EGRB.

In the next section we discuss the evolution of blazar source class. We discuss in section 3 
the luminosity function construction of FSRQs. The methodology to calculate the misaligned blazars contribution 
to the EGRB is given in section 4. We conclude with our discussion in section 5.

\section{Evolution of blazars}
We study the luminosity evolution of blazars by considering a sample that consists of sources from first LAT AGN catalog having $5\sigma$ or above detection significance with $\gamma$-ray flux (E $> 100$ MeV) $>$ $5\times10^{-8}$ ph cm$^{-2}$s$^{-1}$. 
\citet{abdo2010e} showed that Fermi-LAT detects spectrally hard sources at flux level generally fainter 
than those of soft sources. They showed that the LAT sample (including all sources) is complete 
above $7\times10^{-8}$ ph cm$^{-2}$s$^{-1}$ (Figure 1 of \citet{abdo2010e}). However, the maximum photon index 
for blazars is 2.98, much lower than the maximum photon index of all sources of 1st Fermi catalog, 
and from figure 1 of \citet{abdo2010e}, one can see that the blazar sample is complete to a flux 
below $7\times10^{-8}$ ph cm$^{-2}$s$^{-1}$. 
%We consider the limiting flux of our sample as $5\times10^{-8}$ ph cm$^{-2}$s$^{-1}$. 
As a secondary check to the 
completeness of our sample above $5\times10^{-8}$ ph cm$^{-2}$s$^{-1}$, we performed the $<{V}/{V_{max}}>$ test 
of this sample and obtained $<{V}/{V_{max}}>$ that matches within $1\sigma$ of a sample of limiting flux 
$7\times10^{-8}$ ph cm$^{-2}$s$^{-1}$. Our sample consists of $118$ FSRQs. and $39$ BL Lacs ($25$ with known $z$). 
The $<{V}/{V_{max}}>_{\mbox{FSRQ}}$ comes out to be 
$0.63\pm0.03$, which indicates strong evolution. 
For BL Lacs (with known $z$ sources only), $<{V}/{V_{max}}>_{\mbox{BLLac}}$ comes out to be 
$0.49\pm0.06$, which indicates no significant evolution. If we include the sources with unknown $z$ 
considering they are at the average $z$ of the sample, $<{V}/{V_{max}}>_{\mbox{BLLac}}$ comes out to be 
$0.55\pm0.05$. Since, a good fraction of BL Lacs in our 
sample do not have redshift information, it is not possible to construct a meaningful luminosity 
function of BL Lacs at this moment.

We consider both pure luminosity evolution (PLE) and pure density evolution (PDE) for FSRQs. 

%\subsection{pure luminosity evolution}
%We consider the pure luminosity evolution  (density of sources are constant with $z$) of FSRQs. One can write the 
%luminosity of a source at a redshift $z$ as 
%\begin{equation}
%L(z) = L(z=0)\times f(z)
%\end{equation}
\begin{figure}
\centering
\includegraphics[width=80mm]{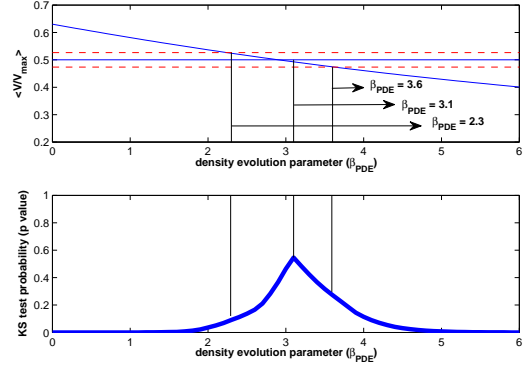}
\caption{Upper panel: The variation of $<\frac{V}{V_{max}}>$ with evolution parameter $\beta$.
Lower Panel: The variation of KS test probability with $\beta$.
} \label{fig-f1}
\end{figure}

Both power law ($(1+z)^{\beta}$) and exponential (exp(${T(z)}/{\tau}$))  evolution function have been considered. 
Here $T(z)$ denotes look-back time. The modified $<{V}/{V_{max}}>$ method is used to find the evolution parameter. 
For the optimum parameter value, ${V}/{V_{max}}$ is expected to be  uniformly distributed between 0 and 1, 
thus $<{V}/{V_{max}}> = 0.5$. 
%Fig. \ref{fig1} shows the variation of $\frac{V}{V_{max}}$ with different $\tau$ values. Horizontal dotted lines show the 1 $\sigma$ error in $\frac{V}{V_{max}}$. 
We find $\tau_{PLE} = 0.21 ^{+0.05}_{-0.02}$.
For each value of evolution parameter ($\tau$), the distribution of ${V}/{V_{max}}$ is compared with a uniform distribution using KS-test which shows distribution of ${V}/{V_{max}}$ is uniform for $\tau_{PLE} = 0.21 ^{+0.05}_{-0.02}$. 
%Fig. \ref{fig2} shows the variation of  $<\frac{V}{V_{max}}>$ with $\beta$ considering the power law evolutionary model. 
The optimum value of  $\beta_{PLE}$ is found to be 
$2.1^{+ 0.2}_{-0.4}$.  
Similarly, for pure density evolution, the parameter values of exponential and power law 
evolution models are, $\tau_{PDE} = 0.17 ^{+0.07}_{-0.03}$ and $\beta_{PDE} = 3.1 ^{+0.5}_{-0.8}$  respectively. The upper panel of Fig \ref{fig-f1} shows the variation of $<{V}/{V_{max}}>$ with different values of power law PDE parameter $\beta_{PDE}$. Horizontal dotted 
lines show the $1 \sigma$ error in $<{V}/{V_{max}}>$. 

We derived the average $\gamma$-ray photon index ($s_{\gamma}$) following the methodology given in 
\citep{venters2007,deb1}. The values of $<s_{\gamma}>$ for FSRQs and BL Lacs are $2.45\pm0.16$ and $2.17\pm0.22$ respectively.

\section{Luminosity function construction of FSRQs}
For pure luminosity evolution, after de-evolution of the sources at $z = 0$, 
we constructed their luminosity function by ${1}/{V_{max}}$ method which 
suggests a luminosity function described by a broken power law. We then 
used maximum Likelihood function for the redshift
distribution of sources, considering a broken power law luminosity function,

\begin{eqnarray}
\phi(L_0) & = &  {\phi}_0 \times \Bigg (\frac{L_0}{L_B}\Bigg )^{-{\alpha}_1} \, \, \, \, L_0 \le L_B \, , \nonumber \\
& = & {\phi}_0 \times \Bigg(\frac{L_0}{L_B}\Bigg )^{-{\alpha}_2} \, \, \, \, L_0 > L_B \, .
\end{eqnarray}
Here $\phi(L_0)$ is the de-evolved luminosity function and $\phi_0$ is the normalization of the luminosity function. 
The break luminosity ($L_B$) and the lower and upper 
end index ($\alpha_1$ and $\alpha_2$ respectively) are constrained by maximizing 
the likelihood function.

For pure density evolution, we consider a similar form of luminosity function and constrained the parameters 
by maximizing the Likelihood function for the redshift distribution of sources.

\begin{figure}
\centering
\includegraphics[width=80mm]{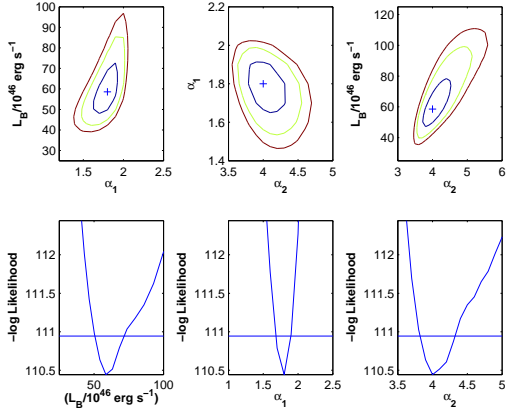}
\caption{Likelihood parameters contour plot of luminosity function for power law PDE model.
Upper Panel: $68.3\%$, $95.4\%$ and $99\%$ ($\Delta log \mathcal{L} = 1.15, 3.08, 4.61$ respectively) 
likelihood contour plot (in to out) 
for $\alpha_1$, $L_B$ (left), $\alpha_1$ (middle), $\alpha_2$ and  $\alpha_2$ (right), $L_B$.
Lower Panel: variation of negative log likelihood values for $L_B$ (left), $\alpha_1$ (middle) and $\alpha_2$ (right). 
the intersections of the solid line with the curve represents the $1 \sigma$ value.} \label{fig-f2}
\end{figure}

The derived luminosity function parameters values are given in Table \ref{Tab:lumfn-param}. 
Fig \ref{fig-f2} shows the contour plots of different luminosity function parameters for power law 
PDE model. 

\begin{figure}
\centering
\includegraphics[width=80mm]{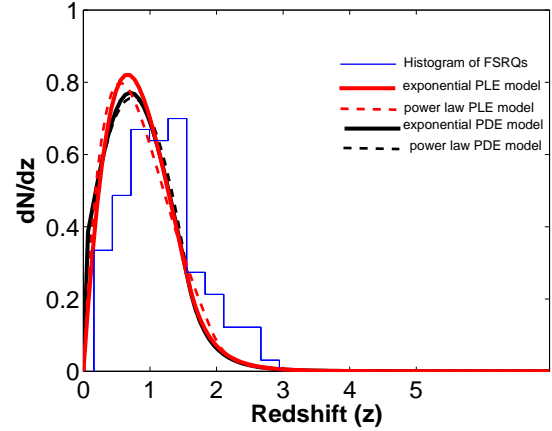}
\caption{Distribution of FSRQs with redshift. Histogram shows the
  Fermi observed distribution. The red curves represent PLE models. 
The black curves represent PDE models. solid and dash curves are
exponential and power law models respectively. } \label{fig-f3}
\end{figure}

The normalization constant of luminosity function has been determined by equating the total 
number of FSRQs in our sample to that expected from the model luminosity function. 
The source distribution ($dN/dz$) has been over plotted with the model distribution 
for PLE and PDE models in Fig \ref{fig-f3}. Exponential and power law evolution models are 
considered for both PLE and PDE models.

\begin{table}[t]
\begin{center}
\caption{Luminosity Function Parameters}
\begin{tabular}{|l|c|c|c|}
\hline
& \textbf{$L_B$ (in $10^{46}$\,erg\,s$^{-1}$)}  & \textbf{$\alpha_1$}  & \textbf{$\alpha_2$} \\
\hline $\tau_{PLE}$  & $2.48^{+0.82}_{-0.33}$ & $1.5\pm 0.3$ & $4.5^{+0.8}_{-0.3}$ \\
\hline $\beta_{PLE}$ & $11.68^{+2.12}_{-1.61}$ & $1.5\pm 0.2$ & $7.1^{+2.4}_{-1.1}$ \\
\hline $\tau_{PDE}$  & $61.93^{+21.68}_{-15.20}$ & $2.0\pm 0.1$ & $3.7^{+0.4}_{-0.2}$ \\
\hline $\beta_{PDE}$ & $59.73^{+11.71}_{-9.02}$ & $1.8 \pm 0.1$  & $4.0^{+0.3}_{-0.2}$\\
\tableline
\end{tabular}
\label{Tab:lumfn-param}
%% Any table notes must follow the \end{tabular} command.
%\tablecomments{We can also attach a long-ish paragraph of explanatory
%material to a table.}
\end{center}
\end{table}

\section{Misaligned blazar contribution}
According to the AGN Unification scenario, FR II galaxies are the parent population of FSRQs whereas, FR I galaxies are the parent population 
of BL Lacs. 
The AGN jet emission is expected to fall very rapidly with increasing jet to line-of-sight angle. Nevertheless, considering the large 
population of these misaligned blazars compared to nearly-aligned ones, these sources could 
contribute significantly to EGRB. Fermi has already detected 11 misaligned blazars.  

The observed jet luminosity $L(\theta)$ at a jet to line-of-sight angle 
($\theta$) can be written as
\begin{equation} 
L(\theta)=L(0)\times \xi(\theta)
\end{equation}
From the knowledge of $\xi(\theta)$ one can in principle, estimate 
the contribution from off-axis AGNs to the EGRB.

\subsection{Construction of $\xi(\theta)$}

Our aim is to model the SED of all Fermi-detected off-axis sources using standard leptonic
models with structured jets. 
From the knowledge of these SED modeling and VLBI observations one can fix the emission 
processes in the jet and other jet parameters and can construct $\xi(\theta)$. 
%using SED model
%and the bulk Lorentz factor ($\Gamma$) distribution of FSRQs and BL Lacs. 
Next step is to constrain $\xi(\theta)$ by comparing $\gamma$-ray detected misaligned blazars with 
equivalent on-axis objects (by matching sources with
similar jet power, $\Gamma$) and then validate our
predictions by modeling few misaligned blazars that are not detected by Fermi.

From the observed $\gamma$-ray luminosities of off-axis sources 
and their jet inclination angle one can also construct $\xi(\theta)$.
The difference in the observed $\gamma$-ray luminosities between two radio galaxies 
can arise from two different reasons:
\newline a$)$ due to their different jet to line-of-sight angle,
\newline b$)$ due to their different intrinsic luminosities.

If one considers that the total radio luminosity is a tracer of unbeamed luminosity, 
then after an appropriate scaling, one can construct $\xi(\theta)$ from scaled 
$\gamma$-ray luminosities and jet inclination angles of off-axis sources. 

\begin{figure}
\centering
\includegraphics[width=80mm]{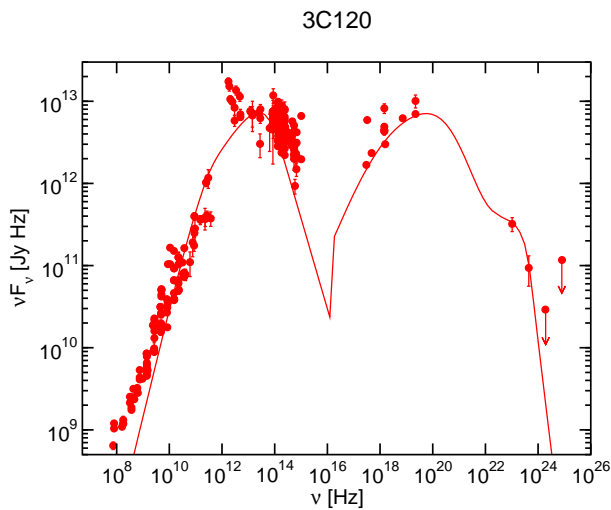}
\caption{SED modeling of misaligned blazars 3C 120} \label{fig-f4}
\end{figure}

We have initiated this work. Our aim is to analyze Fermi data to date on radio
galaxies and model it to derive jet parameters. We have completed the analysis and the SED 
modeling of 3C111 and 3C120 (Figure \ref{fig-f4} shows the SED fitting
of 3C 120.)

%\centerline{
%\begin{figure*}[t]
%\centering
%\includegraphics[width=80mm]{/media/DEB-8GB/LUMFERMI18112011/3C111_1.eps}
%\caption{SED modeling of misaligned blazars 3C 111} \label{fig-f2}
%\end{figure*}

Both the sources are well fitted with simple SSC emission model. 
$\Gamma$ is also $\sim$5 for both the sources.

\section{Discussion}
Using first Fermi catalog, we investigated the luminosity function and nature of evolution of FSRQs and BL Lacs 
separately in $\gamma$-rays. The nature of evolution of FSRQs and BL Lacs as found in this work, 
are in agreement with earlier works (e.g., \cite{deb1,chiang1998}). We considered both PLE and PDE models for FSRQs. 
Both power law and exponential evolution functions have been considered. The luminosity function 
is considered to be of a broken power law form and the parameters are found using likelihood analysis. 
From this analysis it is not possible to strongly conclude on the true evolution model. Fig \ref{fig-f3} suggests that PDE model 
explains observations marginally better than that of PLE model. 
In contrast to earlier works of determination of 
luminosity function from $\gamma$-ray observations 
\citep{deb1,chiang1995,chiang1998}, our derived luminosity function is well constrained. 
%From the derived luminosity function we determined their contribution 
%to the EGRB. Our estimated value is in agreement with that found by 
%\citet{abdo2010e}. 
In a recent work, Ajello et al.~\cite{ajello2011} showed that the redshift of the peak in the number density of FSRQs is 
luminosity dependent. Hence, they considered luminosity dependent density evolution for 
FSRQs (similar to that of radio-quiet AGN). They considered a much lower 
limiting flux ($\sim 10^{-8}$ ph cm$^{-2}$ s$^{-1}$) and hence, a much larger 
sample ($186$ FSRQs) than us. 

For misaligned blazars, we plan for SED fitting of all detected off-axis sources 
and then construct $\xi(\theta)$ from model. This work is under progress and will be reported elsewhere.

\begin{acknowledgments}
We are thankful to Dipankar Bhattacharya, IUCAA, for his valuable 
comments and suggestions. 

This work is partially supported by the 
NASA grant NNX09AT71G.
\end{acknowledgments}

\bigskip % extra skip inserted

\begin{thebibliography}{9}   % Use for  1-9  references
%\begin{thebibliography}{99} % Use for 10-99 references
%\bibitem{accelconf-ref}
%http://www.cern.ch/accelconf
%\bibitem[Kraushaar et al.(1972)]{kraus2} Kraushaar W.L., Clark G. W., Garmire G. P., Borken R., Higbie P., Leong V., Thorsos T., 1972, ApJ, 177, 341.
%\bibitem[Thompson \& Fichtel(1982)]{thompson1982} Thompson D.J., \& Fichtel C.E., 1982, A\&A, 109, 352.
%\bibitem[Kanbach et al.(1988)]{kanbach1988} Kanbach, G. et al. 1988, Space Sci Rev., 49, 69.
%\bibitem[Sreekumar et al.(1998)]{sreekumar1998} Sreekumar P. et al., 1998, ApJ, 494, 523. 
%\bibitem[Strong, Moskalenko \& Reimer(2004b)]{strong2004b} Strong A.W., Moskalenko I.V., Reimer O., 2004b, ApJ, 613, 956.
%\bibitem[Strong, Moskalenko \& Reimer(2004a)]{strong2004a} Strong A.W., Moskalenko I.V., Reimer O., 2004a, ApJ, 613, 962. 
\bibitem[Cohen, R$\acute{u}$jula \& Glashow(1998)]{cohenrujulaglashow1998} Cohen A.G., R$\acute{u}$jula A.D., Glashow S.L., 1998, ApJ, 495, 539.
\bibitem[Gabici \& Blasi(2003)]{gabici2003} Gabici S., Blasi P., 2003, 
APh.., 19, 679.
\bibitem[Loeb \& Waxman(2000)]{loeb2000} Loeb A., Waxman E., 2000, Nature., 405, 156.
\bibitem[Abdo et al.(2010b)]{abdo2010b} Abdo, A. A. et al. 2010b, PRL, 104, 101101.
\bibitem[Abdo et al.(2010c)]{abdo2010c} Abdo, A. A. et al., 2010c, ApJS, 188, 405.
\bibitem[Abdo et al.(2010a)]{abdo2010a} Abdo, A. A. et al., 2010a, ApJ, 715, 429.
\bibitem[Abdo et al.(2010d)]{abdo2010d} Abdo A. A. et al., 2010d, ApJ, 720, 912.
\bibitem[Abazajian, Blanchet \& Harding(2010)]{abazajian2010} Abazajian, K.N.; Blanchet, S., \& Harding, J. P., 2010, (arXiv: 1012.1247).
\bibitem[Cavadini, Salvaterra \& Haardt(2011)]{cavadini2011} Cavadini M., Salvaterra R., \& Haardt F., 2011, (arXiv 1105.4613).
\bibitem[Inoue(2011)]{inoue2011} Inoue, Y., 2011, ApJ, 733, 66.
\bibitem[Li \& Cao(2011)]{li2011} Li, F., \& Cao, X., 2011, (arXiv: 1103.4545).
\bibitem[Narumoto \& Totani(2006)]{narumoto1} Narumoto T., Totani T., 2006, ApJ, 643, 81.
\bibitem[Narumoto \& Totani(2007)]{narumoto2} Narumoto T., Totani T., 2007, Ap\&SS, 309 73.  
%\bibitem[Sowards-Emmerd, Romani \& Michelson(2003)]{sowards2003} Sowards-Emmerd D., Romani R.W., Michelson P.F., 2003,
% ApJ, 590, 109.
%\bibitem[Sowards-Emmerd et al.(2004)]{sowards2004} Sowards-Emmerd D., Romani R.W., Michelson P.F., Ulvestad J.S., 2004, ApJ, 609, 564.
\bibitem[Stecker, Salamon \& Malkan(1993)]{stecker1993} Stecker F.W., Salamon M.H., Malkan M.A., 1993, ApJ, 410, L71.
\bibitem[Stecker \& Salamon(1996)]{stecker1996} Stecker F.W., Salamon M.H., 1996, ApJ, 464, 600.
\bibitem[Stecker \& Venters(2011)]{stecker2011} Stecker, F.W., Venters, T.M., 2011, ApJ, 736, 40.
\bibitem[Bhattacharya, Sreekumar \& Mukherjee(2009a)]{deb1} Bhattacharya D., Sreekumar P., Mukherjee R., 2009, RAA, 9, 85.
\bibitem[Bhattacharya, Sreekumar \& Mukherjee(2009b)]{deb2} Bhattacharya D., Sreekumar P., Mukherjee R., 2009, RAA, 9, 1205.
\bibitem[Chiang \& Mukherjee(1998)]{chiang1998} Chiang J., Mukherjee R., 1998, ApJ, 496, 752.
\bibitem[Chiang et al.(1995)]{chiang1995} Chiang J. et al., 1995 ApJ, 452, 156.
\bibitem[M$\ddot{u}$cke \& Pohl(2000)]{mucke} M$\ddot{u}$cke A., Pohl M., 2000, MNRAS, 312, 177.
\bibitem[Dermer(2007)]{dermer2007} Dermer C.D., 2007, ApJ, 659, 958.
\bibitem{abdo2009a} Abdo, A. A. et al., 2009, ApJ, 700, 597.
\bibitem[Abdo et al.(2010e)]{abdo2010e} Abdo, A. A. et al., 2010e, ApJ, 720, 435.
\bibitem[Venters \& Pavlidou(2007)]{venters2007} Venters T.M., Pavlidou V., 2007, ApJ, 666, 128.
\bibitem[Ajello et al.(2011)]{ajello2011} Ajello, M. et al., 2011, (arXiv:1110.3787v1)
%\bibitem[Hartman et al.(1999)]{hartman1999} Hartman R.C. et al., 1999, Ap\&SS, 123, 79.
%\bibitem{exampl-ref}
%A.N. Other, ``A Very Interesting Paper'', EPAC'96, Sitges, June
%1996.

%\bibitem{templates-ref}
%http://www.cern.ch/accelconf/templates.html

\end{thebibliography}

\end{document}